\documentstyle[epsf,epsfig,rotating]{article}

\input epsf
\begin{document}

\begin{center}

{\large \bf Is Strangeness still interesting at RHIC ?}

\bigskip

\end{center}

\begin{center}
Rene Bellwied
\end{center}

\begin{center}
Wayne State University, Department of Physics and Astronomy,\\ 
Detroit, MI 48202, U.S.A., E-mail: bellwied@physics.wayne.edu\\
\end{center}

\bigskip

\bigskip

\begin{abstract}

With the advent of the Relativistic Heavy Ion Collider (RHIC) at
Brookhaven National Laboratory (BNL), Heavy Ion Physics will enter 
a new energy regime. The question is whether the signatures proposed 
for the discovery of a phase transition from hadronic matter to a 
Quark Gluon Plasma (QGP), that were established on the basis of
collisions at the BEVALAC, the AGS, and the SPS, respectively, are
still useful and detectable at these high incident energies. In the
past two decades, measurements related to strangeness formation in
the collision were advocated as potential signatures and were tested
in numerous fixed target experiments at the AGS and the SPS. In this
article I will review the capabilities of the RHIC detectors to measure
various aspects of strangeness, and I will try to answer the question 
whether the information content of those measurements is comparable to 
the one at lower energies.

\end{abstract}

\section{Introduction}

Since the beginning of Relativistic Heavy Ion Collisions at the 
BEVALAC, the quest for evidence of a phase transition between hadronic
matter and a chirally symmetric and deconfined phase, called the
Quark Gluon Plasma, has led to numerous proposed hadronic and leptonic
signatures. Early theoretical suggestions of hadronic signatures had to be 
revised after it was shown that final state interactions affect the 
initial formation observables of hadrons. One potential signal that is still
being considered is strangeness formation. Measurements relating to 
strangeness production
in a deconfined phase are numerous and span from simple strangeness yields
(as evidence for strangeness enhancement), over strangeness ratios (as
evidence for strangeness equilibration) all the way to strangeness distillation
(as evidence for chiral symmetry restoration or strange quark matter
formation). In the following sections, I will show that
the RHIC detectors are well equipped to repeat and extend the strangeness
measurements performed at the AGS and SPS, and I will give you a personal
view on the significance of those measurements at RHIC based on some
recent theoretical work.
 
I will start by briefly describing the accelerator and
the various RHIC experiments. I will then focus on the STAR detector, which
addresses most strangeness measurements at RHIC. After a short
description of the relevant detector components, I will present detector
performance simulations relevant for strangeness measurements. 
All simulations shown were provided by the various RHIC detector
groups and should be considered preliminary. 

In the Discussion
and Conclusions chapter I will relate those measurements to some recent
theoretical work and then draw my conclusion on whether strangeness is
still interesting at RHIC. The reader should complement this article
with many of the articles in these proceedings, which essentially show
that the measurement of strangeness ratios at the AGS and SPS might have 
led to a deep understanding of the collision dynamics and possible evidence
for transition like behavior at the SPS.

We do not know what nature
has in mind for us at RHIC, and therefore any model description has to
be incomplete. Still, even on the basis of the most basic models, we can
establish a significant change in the collision environment from
SPS to RHIC. This change is documented in the above table which shows
a comparison between
SPS and RHIC parameters, as presented by Dumitru and Rischke
\cite{risch98}, under certain model assumption, namely a parton cascade
with a hydrodynamical 'afterburner' to describe the hadronization phase.
In this case the system follows a cylindrically symmetric, longitudinally
boost invariant expansion. Clearly, the enhancements in the entropy per
baryon and the energy density are large, whereas the baryon density
shows the expected decrease. Because the plasma in the central region, 
if formed, will most likely
be baryon-poor at these energies, certain strangeness physics observables
which rely on high baryon density, like strange quark matter formation
\cite{carsten1} or medium modifications \cite{shury1}, \cite{mueller1}, might 
be less
likely to occur at mid-rapidity, where most detectors have their best 
coverage. Still, measurements in the forward regions might
yield access to a more baryon rich plasma.

\vspace{0.2in}
\begin{table}
\centering
\begin{tabular} {|c|c|c|} \hline
{\bf Parameter} & {\bf RHIC} & {\bf SPS} \\ \hline
Energy density $\epsilon_{i}$ & 17 GeV/fm$^{3}$ & 5.3 GeV/fm$^{3}$ \\ \hline 
Baryon Density $\rho_{B}$ & 2.3 $\rho_{0}$ & 4.5 $\rho_{0}$ \\ \hline
dN$_{B}$/dy & 25 & 80 \\ \hline
Initial Temperature T$_{i}$ & 300 MeV & 216 MeV \\ \hline
Chemical Potential $\mu_{q}$ & 47 MeV & 167 MeV \\ \hline
Strange chem. potential $\mu_{s}$ & 0 MeV & 0 MeV \\ \hline
Entropy per baryon S/$\rho_{B}$ & 200 & 40 \\ \hline
Freeze-out Temperature & 160 MeV & 130 MeV  \\ \hline
Hadronization Time & 16 fm & 10 fm  \\ \hline
\end{tabular}
\end{table}

\section{RHIC and its detectors}

The Relativistic Heavy Ion Collider (RHIC) at BNL will 
enable us for the first time to accelerate heavy ions in collider mode. 
Therefore, it will greatly enhance the energy available in each collision
compared to previous fixed target experiments. The accelerator is scheduled 
to commence data runs by November 1999, in about one year from now. The
maximum available beam energy for Gold Ions will be 100 + 100 A GeV.
The minimum available energy in collider mode will be 30 + 30 A GeV
($\sqrt{s}$ = 60 GeV) due to certain injection constraints. 
This minimum energy might
become important if we try to connect the RHIC measurements to the highest
energy fixed target measurements (the SPS E$_{lab}$ = 160 A GeV corresponds
to a $\sqrt{s}$ = 17 GeV) for the purpose of a signal excitation
function. RHIC is also capable of colliding pp-, pA- and many
other AA-systems. pp-running is scheduled to commence in Year-2 and it
will be complemented by an elaborate Spin Program, based on the acceleration
of polarized protons.

Presently, four detectors are being built for RHIC: two large detectors,
PHENIX \cite{phenix} and STAR \cite{star}, and two smaller devices,
PHOBOS \cite{phobos} and BRAHMS \cite{brahms}. The smaller detectors
are specialized setups with a dedicated program, whereas the
large detectors attempt to measure as many observables in a single collision
as possible. PHENIX has a focus on leptonic signals, with some significant
capabilities in hadron measurements; STAR is dedicated to hadronic
measurements with some limited lepton measurement capabilities. 
In the following I will provide a brief overview of each detector and its
respective strangeness capabilities.

PHOBOS is a table-top two arm spectrometer consisting of various types of
high resolution Silicon detectors. Its main physics goal is to study inclusive
hadron production down to very low transverse momentum. Its strangeness
capabilities include the very precise measurement of neutral and charged
kaon spectra and their respective particle correlations, plus the $\phi$-meson 
decay into the kaon channel.

BRAHMS is a small acceptance spectrometer with variable angle setting, which
is build on the same principle as the successful series of AGS
experiments E802, E859, E866, E902. This detector will yield inclusive
measurements of hadronic particle production over the full rapidity range.
Its main strangeness capabilities include the measurement of charged kaon
spectra, K$^{+}$K$^{-}$-interferometry, plus the $\phi$-meson decay into 
the kaon channel.

PHENIX is a large detector consisting of an axial field magnet with
a two-arm central detector plus muon detection systems attached in each
forward direction. The detector is dedicated to leptonic probes, but it has
very good hadronic detection capabilities in the central arms. This allows
PHENIX not only to measure the kaon spectra very precisely and to very high
transverse momentum, but also the simultaneous measurement of
both $\phi$-meson decay channels, the di-lepton channel and the K$^{+}$K$^{-}$ channel.

STAR is a close to 4$\pi$ coverage detector consisting of a central solenoidal
axially symmetric magnet which hosts the central Time Projection Chamber (TPC),
a Silicon Vertex Tracker (SVT) plus an Electro-Magnetic Calorimeter (EMC).
In the forward directions the coverage is extended by two Forward Time 
Projection
Chambers (FTPC), one on each side. This complete coverage allows event-by-event
analysis of hadronic signals and jets. STAR is the dedicated hadron detector
at RHIC. The crucial detector component for its strangeness measurements is the
Silicon Vertex Tracker, a three barrel vertexing and tracking device,
which enables the reliable reconstruction of secondary and tertiary
vertices for the fast decaying strange mesons and baryons. The SVT
is based on a new Silicon detector technology, called Silicon Drift
Detectors \cite{rehak1}, \cite{rene2}, which was recently successfully
implemented in a 15 plane tracking device in the fixed target heavy ion
experiment E896 at the AGS. This STAR test detector will yield strange 
and multi-strange baryon measurements in the Au+Au system at 11.6 A GeV at
the AGS within the next year.
 
The STAR measurement capabilities include all major strangeness signals, in 
particular the measurements of spectra, ratios and cross sections for K$^{+}$,
K$^{-}$, K$^{0}_{s}$, $\Lambda$, $\overline{\Lambda}$, $\Xi^{-}$, 
$\overline{\Xi^{-}}$, $\Omega^{-}$, and
$\overline{\Omega^{-}}$, in conjunction with $\pi$,p,d etc.
Based on the multiplicities and the detector coverage, only the charged
kaon spectra and the K$^{\pm}$/$\pi^{\pm}$ ratio can be studied event-by-event. 
STAR also measures particle correlations for charged and neutral 
kaons, and possibly $\Lambda\Lambda$-Interferometry. In addition, it 
complements the $\phi$-meson studies by the other three RHIC detectors. 
Finally it enables the search for strange quark matter in form of the 
H-Dibaryon or higher mass strangelets.

Based on the comparison of all the capabilities listed above, I conclude
that there is some limited overlap between detectors, in particular in the 
measurements of the kaon spectrum and the $\phi$-meson decay. The phase space 
coverage and the resolutions are quite different, though, therefore at RHIC,
in contrast to the AGS and SPS experiments, we do not expect many 
different measurements of the same observable. The premier strangeness
detector will be STAR, although PHENIX has an advantage in the basic meson
spectra due to very good particle identification extending to rather large
momenta. STAR has a focus on good rapidity coverage and presently relies, for
the particle identification, on de/dx measurements in the tracking
detectors. For the future, STAR is considering a particle identification 
upgrade either via a RICH detector or a Time of Flight wall, both of which 
might be available at RHIC start-up time.

In the following section I will focus on simulations by my own collaboration,
STAR, and a few simulations graciously provided by PHENIX.

\section{RHIC strangeness simulations}

\subsection{Coverage, Acceptances, and Efficiencies}

Each RHIC detector has different phase-space coverage. Most of the
strangeness measurements will be performed at mid-rapidity in a
pseudo-rapidity window of $\eta$ = $\pm$ 1.5. As a typical example, Fig.1a 
shows the 
geometrical acceptance for K$^{o}_{s}$ and $\Lambda$ particles in the 
central STAR tracking system. Certain measurements with the BRAHMS 
detector and the FTPC in STAR will extend this coverage to forward rapidities.

The transverse momentum coverage is determined by the method applied to
particle identification. In the baseline STAR detector (SVT+TPC) the
particles are identified via energy loss in the various layers of the tracking
detectors. This method allows kaon over pion separation
from the lowest momenta (around 70 MeV/c) to around 700 MeV/c. With the
peak of the kaon spectrum close to 500 MeV/c, particle identification via
de/dx is limited. Higher momentum separation requires more sophisticated
detectors, like a time of flight system (TOF), a ring imaging Cherenkov 
counter (RICH) or a transition radiation detector (TRD). PHENIX, BRAHMS, and
PHOBOS
employ several of those techniques, thus extending the kaon over pion
separation limit to around 1.5 GeV/c.

Reconstruction efficiencies are crucial for the detection of decaying strange
mesons and baryons. The efficiency might depend on the number of initial 
hadrons. Parton Cascades with realistic hadronization scenarios 
(e.g. VNI, HIJING-B) presently yield a significantly higher particle 
multiplicity than standard string fragmentation models (e.g. FRITIOF, 
VENUS 4.12). 
Performance simulations of the STAR tracking system, though, show no 
degradation in reconstruction efficiencies up to initial particle 
multiplicities of four times the yield of a string fragmentation model. 
The standard numbers for
particle occupancies presently being used are 1000 charged pions,
150 charged kaons, and around 150 baryons per unit rapidity. Within the 
coverage of the central detector of STAR we thus expect a total of about 2000 
charged particles per event.

Fig.1b shows tracking efficiencies for secondary particles (decay products) 
in STAR as a function of the transverse momentum of the particle.

\begin{figure}[hbtl]
\begin{center}
\epsfig{file=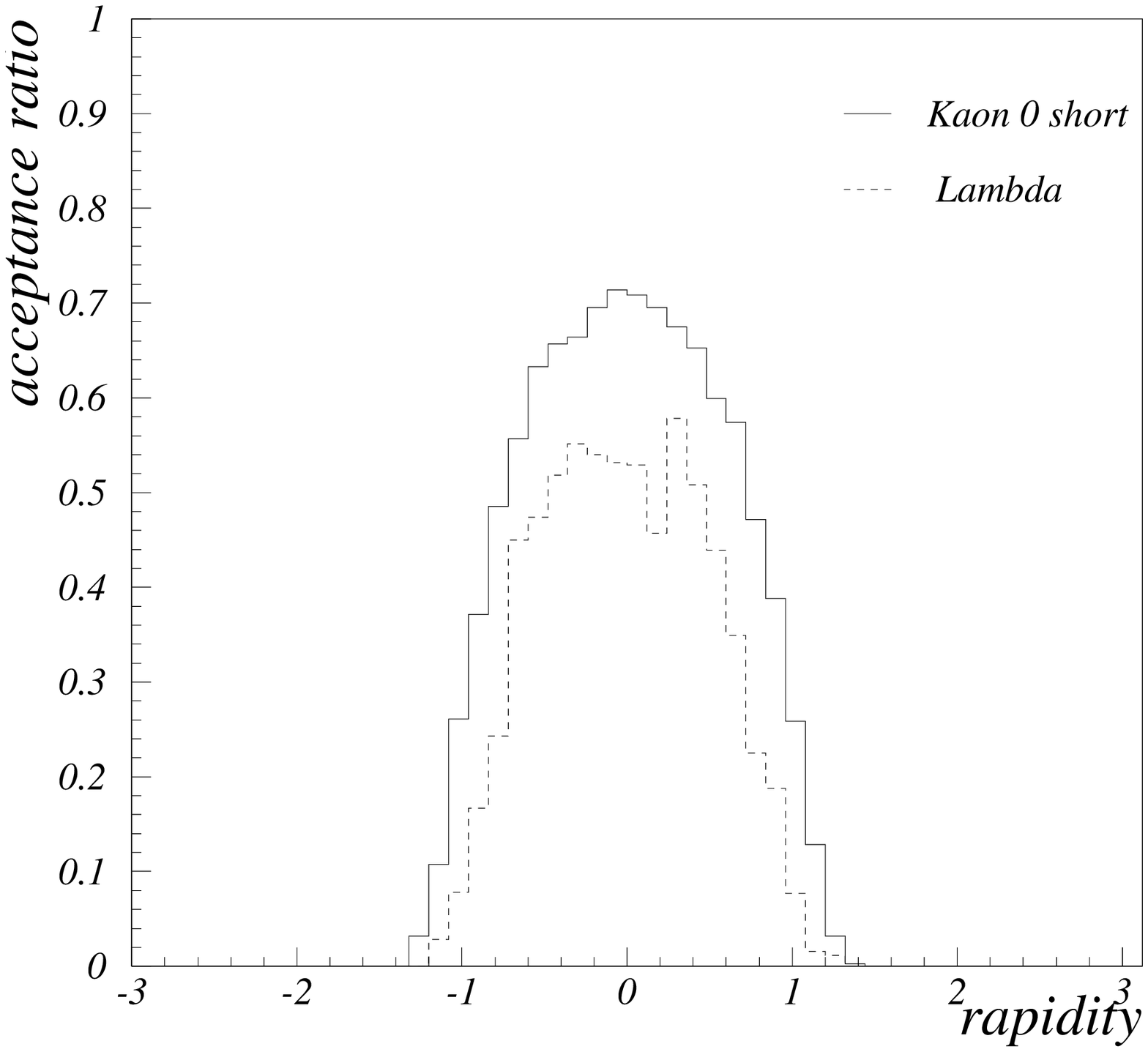,bbllx=100pt,bblly=155pt,bburx=520pt,bbury=645pt,width=6.cm}
\epsfig{file=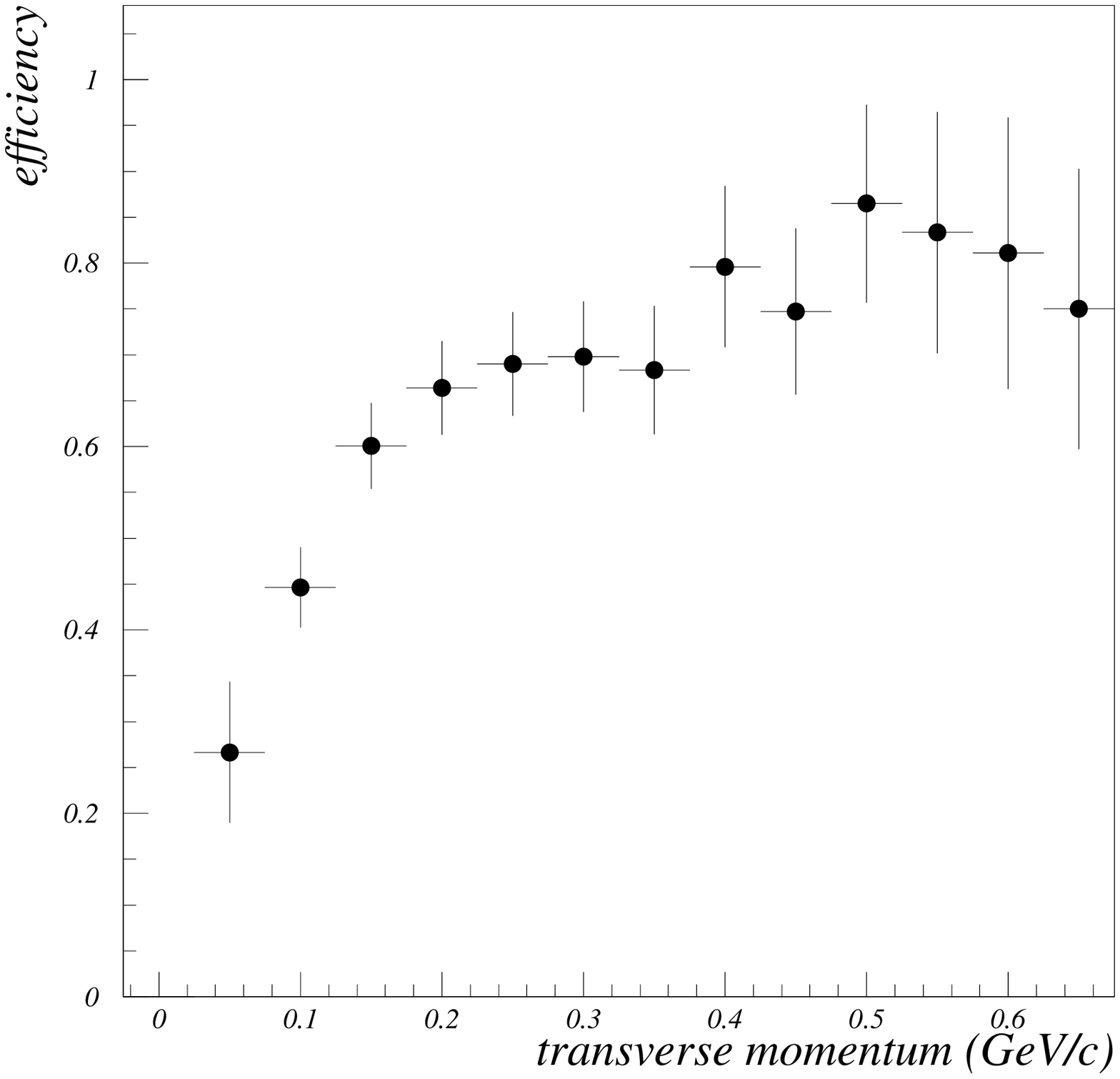,bbllx=0pt,bblly=155pt,bburx=520pt,bbury=645pt,width=7cm}
\end{center}
\caption{ 
a.) Geometrical acceptance of K$^{0}_{s}$ and $\Lambda$ 
in the integrated STAR tracking system
b.) Secondary track reconstruction efficiency for SVT alone as a 
function of momentum.
}
\end{figure}

\subsection{Kaon capabilities}

At RHIC, kaons will still account for close to 90$\%$ of the 
strangeness yield, based on standard string fragmentation models. 
Strange baryons, though very interesting as rare probes of strangeness 
equilibration and in-medium behavior, will contribute only moderately to 
the strangeness yield.

The most basic strangeness quantity, the charged kaon spectrum,
will be measured by all four detectors. PHENIX has very precise
particle identification over a large momentum range and will therefore yield 
a reliable inclusive measurement of the transverse momentum spectrum.
The other three experiments will be able to contribute to the charged kaon 
spectrum at different levels. PHOBOS will reduce the transverse momentum lower 
limit from the PHENIX cut-off at 400 MeV/c down to below 100 MeV/c. BRAHMS 
will measure the kaon spectrum at various rapidity settings away from the 
central coverage of PHENIX and PHOBOS. And STAR will measure the kaon 
spectrum in a limited momentum range, but its 2$\pi$ coverage allows it 
to measure the kaon yield in this momentum range on an event-by-event basis.
Fig.2 compares different detection methods and their respective momentum 
range for protons, pions, and kaons in STAR. The STAR-SVT/TPC de/dx measurement 
covers the kaon spectrum from around 70 MeV/c to around 600 MeV/c. The
proposed RICH upgrade will identify kaons for around 1.2 GeV/c to 3 GeV/c.
The PHENIX TOF wall will cover the kaon spectrum from 400 MeV/c to about 
1.5 GeV/c.

\begin{figure}[htbl]
\begin{center}
\mbox{\epsfig{file=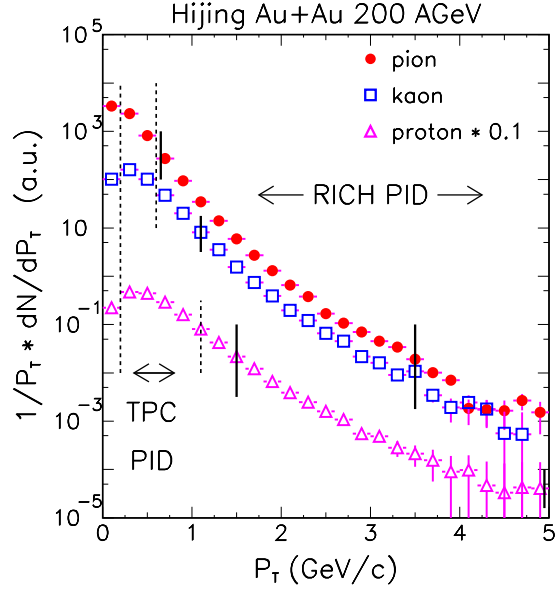,bbllx=0pt,bblly=0pt,bburx=460pt,bbury=480pt,width=8cm}}
\end{center}
\caption{Transverse momentum spectrum of charged protons, pions, and kaons, and
the ranges of several particle identification devices}
\end{figure}

Based on the kaon yield per event, recent STAR simulations showed that
the K/$\pi$ ratio can be determined event by event for a certain momentum range 
(p$_{T}$= 200 - 600 MeV/c), in which the particle identification via de/dx 
yields reliable results. The error on those measurements is on the order
of 10$\%$. Fluctuations in the event-by-event K/$\pi$ ratio might help in
selecting a set of events that show an unusually high level of strangeness
enhancement.

The large number of kaons also allows very precise charged kaon-interferometry 
measurements in all four experiments. As in previous fixed target experiments 
at the AGS and SPS, two particle interferometry
will be used to extract the source size of the system at freeze-out.
In addition, PHOBOS and STAR have good coverage for measuring the K$^{0}_{s}$ yield
and spectrum, and STAR is unique in measuring K$^{0}_{s}$-K$^{0}_{s}$
correlations . Neutral particle interferometry measurements add 
several advantages to the HBT analysis \cite{opal}. Furthermore, it
was suggested by Greiner and Mueller , that both K$^{0}_{s}$ 
and $\Lambda$ interferometry could provide potential QGP formation 
signatures \cite{carsten2}.


\subsection{Strange and Multi-Strange Baryons}

The detection of $\Lambda$, $\Xi$, and $\Omega$ particles is a unique
feature of STAR. Fig.3a shows an Armenteros plot demonstrating K$^{0}_{s}$ and 
$\Lambda$ separation based on the combined tracking system information. 
Figs.3b show the present status 
of invariant mass reconstruction efforts for K$^{0}_{s}$, $\Lambda$ and 
$\Xi^{-}$. A good mass resolution and a good signal to noise ratio for all
strange particles is obtained. 
The efficiency for the $\Omega^{-}$ reconstruction, which is not shown here,
is expected to be comparable to that of the $\Xi^{-}$ reconstruction, 
see Table 2.

\begin{figure}[hbtl]
\begin{center}
\mbox{
\epsfig{file=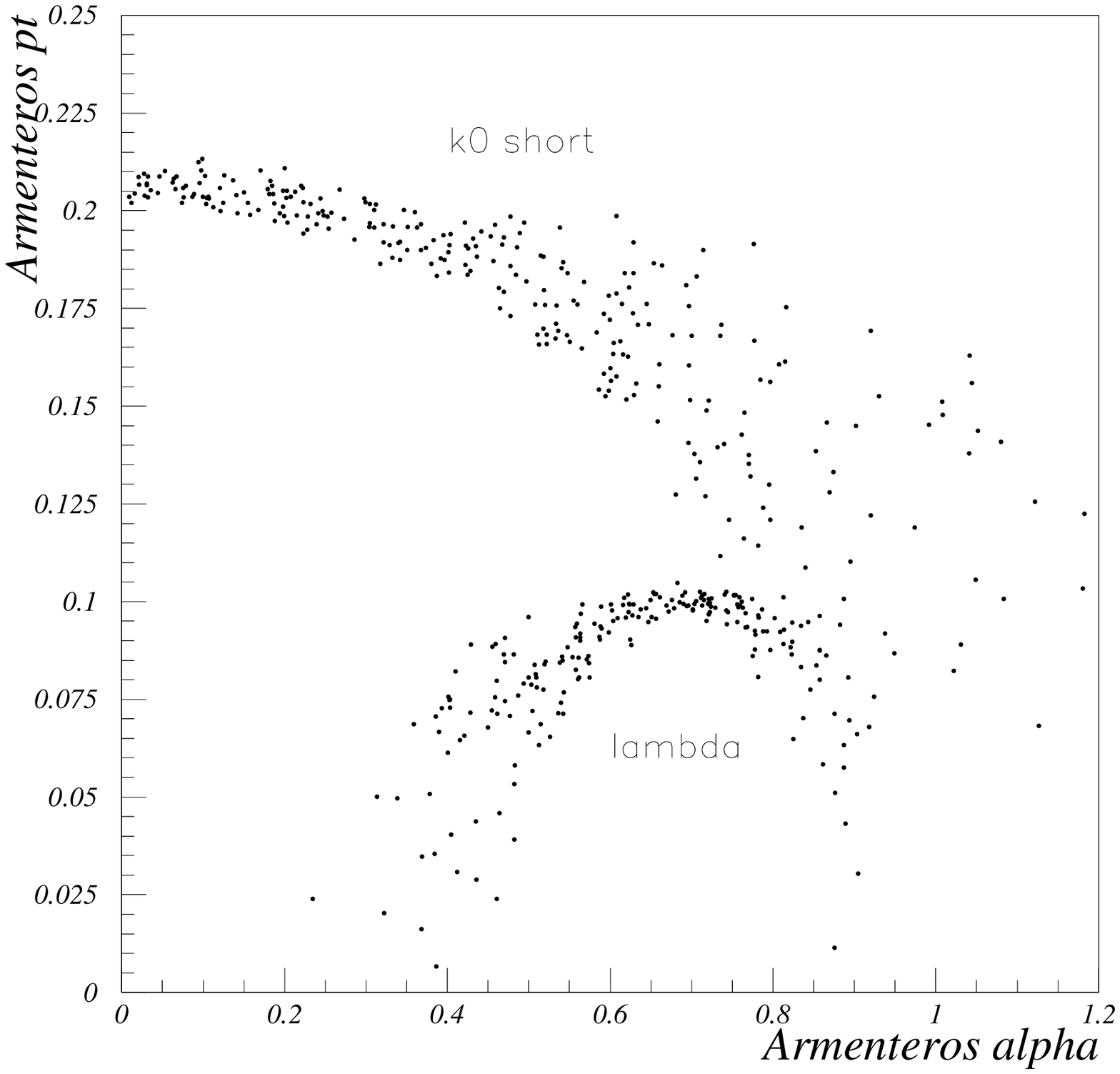,
bbllx=50pt,bblly=155pt,bburx=520pt,bbury=645pt,width=5.5cm}
}
\mbox{
\epsfig{file=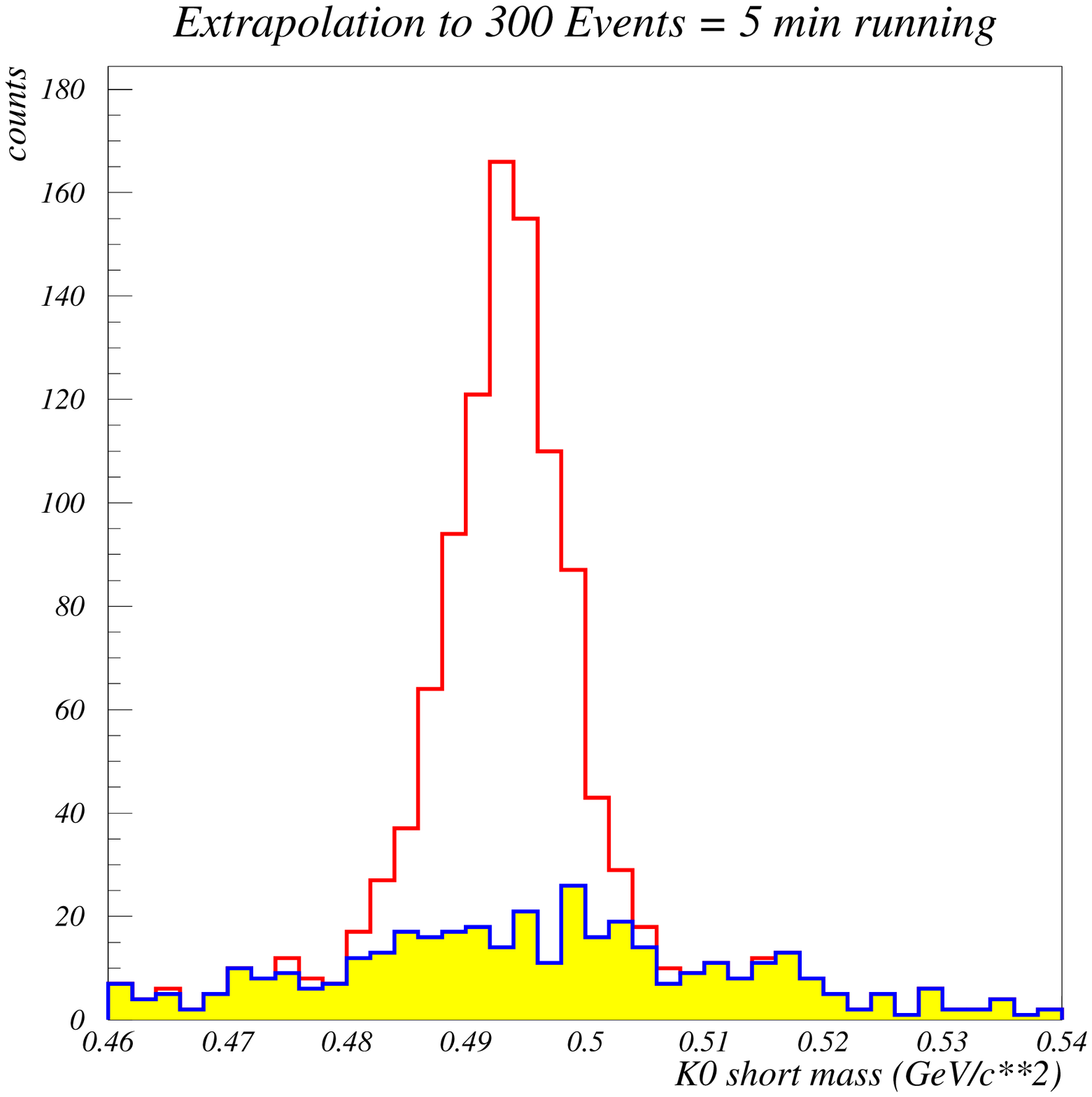,bbllx=50pt,bblly=155pt,bburx=520pt,bbury=645pt,width=5.5cm}
}
\mbox{
\epsfig{file=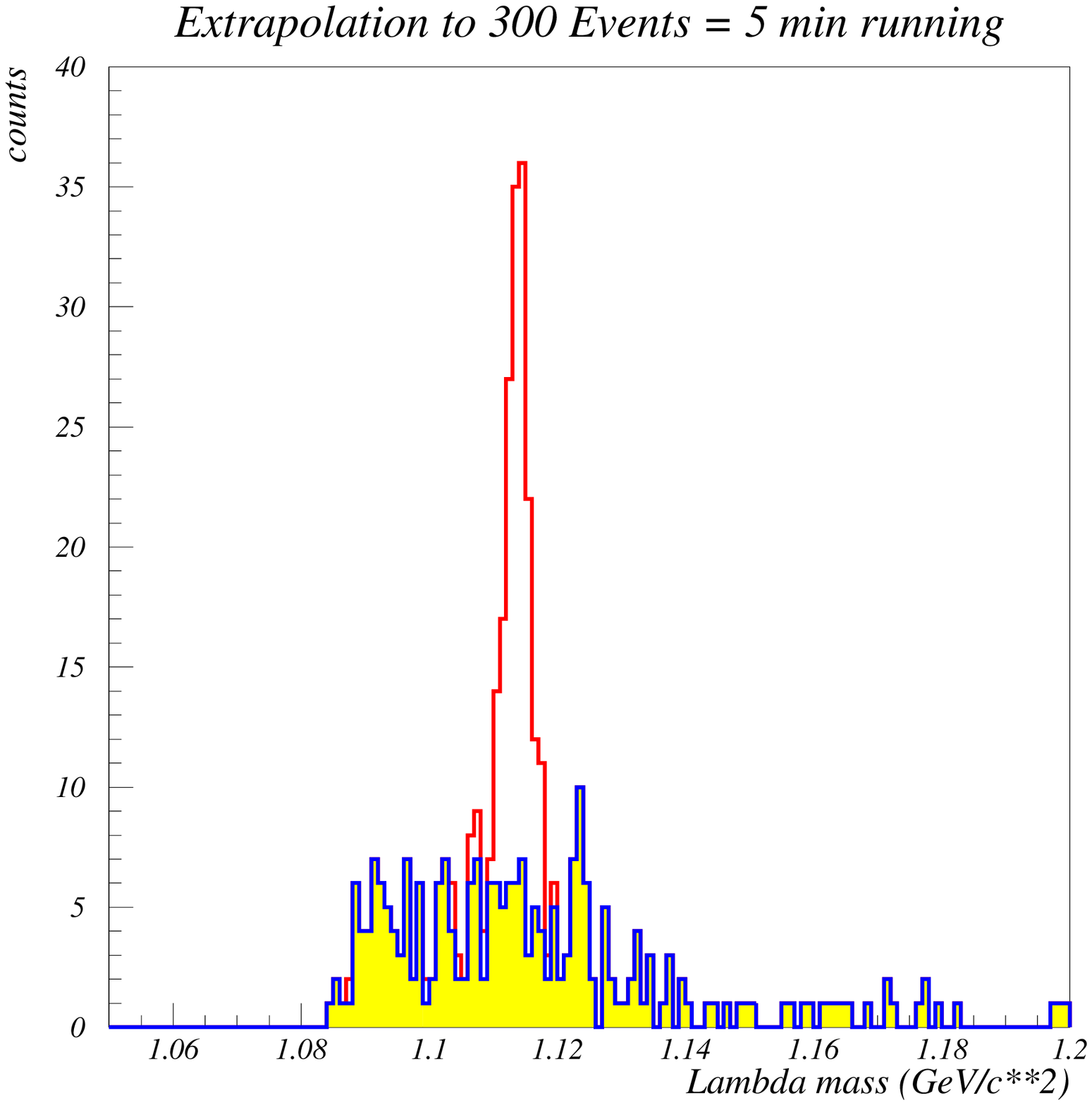,bbllx=50pt,bblly=155pt,bburx=520pt,bbury=700pt,width=5.5cm}
}
\mbox{
\epsfig{file=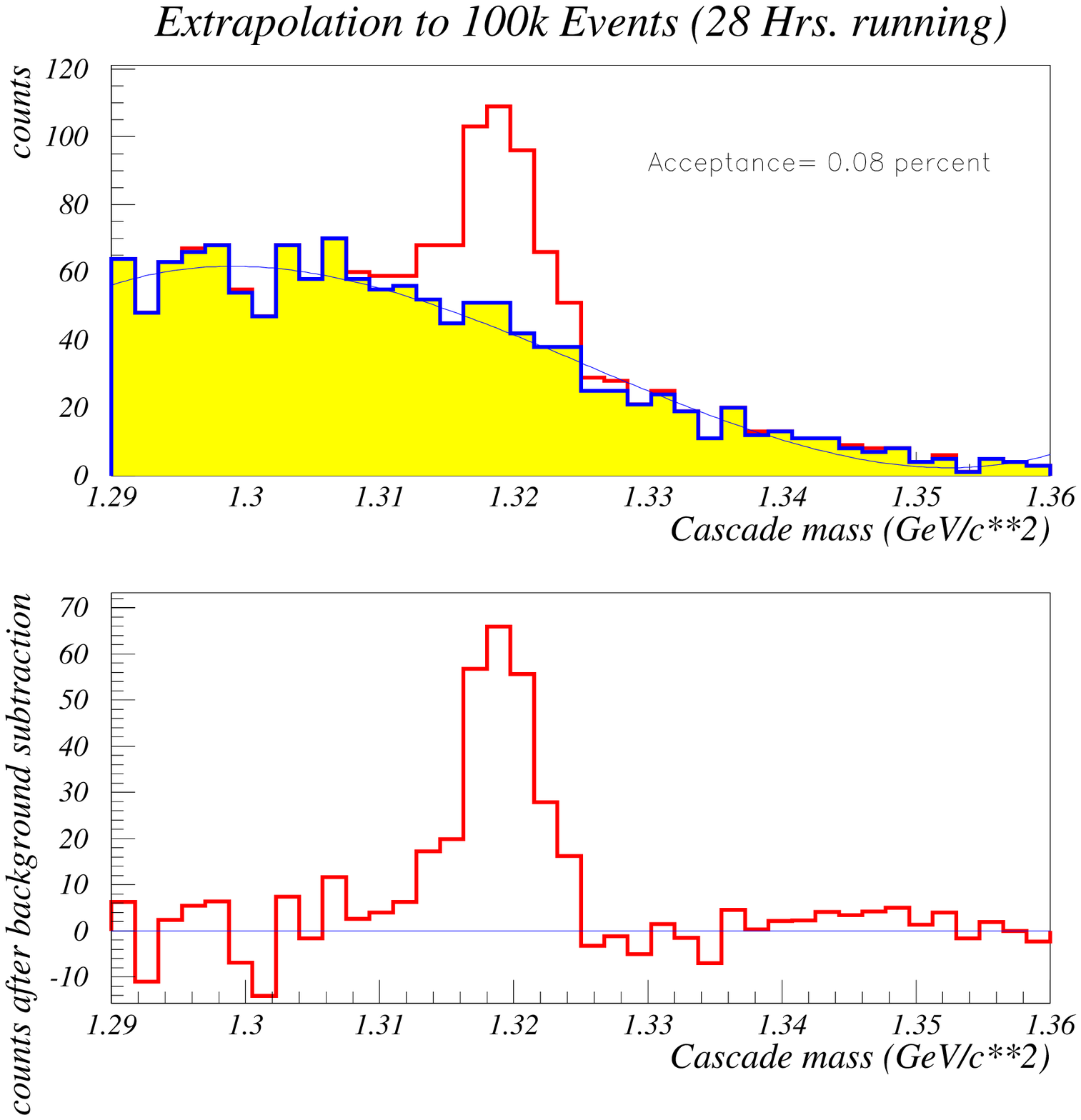,bbllx=50pt,bblly=155pt,bburx=520pt,bbury=700pt,width=5.5cm}
}
\end{center}
\caption{a.) Armenteros plot showing $K_s^0$ and $\Lambda$ separation based on SVT information
 b.) Invariant mass spectra of reconstructed K$^{0}_{s}$,$\Lambda$, and 
$\Xi^{-}$ particles based on SVT and TPC tracking information.}
\end{figure}

\clearpage

Table 2 shows the yield per event and the estimated run time to 
obtain a significant sample of strange particles in STAR.

\vspace{0.2in}
\begin{table}
\centering
\begin{tabular} {|c|c|c|c|} \hline
& &  \\
& {\bf Efficiency} &  & {\bf run time} \\
{\bf Particle} & {\bf incl. acceptance and} & {\bf particles/event} &  {\bf to obtain} \\
& {\bf reconstruction efficiency} & & {\bf 1000 particles} \\ \hline
K$^{o}$ & 3.7$\%$ & 2.5 & 7 min. \\ \hline
$\Lambda$ & 8$\%$ & 0.5 & 35 min. \\ \hline
$\Xi^{-}$ & 0.08$\%$ & 0.003 & 4.1 days \\ \hline
$\Omega^{-}$ & 0.07$\%$ & 0.00005 & several months \\ \hline
\end{tabular}
\end{table}

From this table
we conclude that a.) no strangeness signal besides the charged kaon spectrum 
can be measured event-by-event, and b.) it will take around six months to
collect the required statistics for a good multistrange baryon measurement.
It is interesting to point out that ALICE at the LHC will be able to
measure at least the K$^{0}_{s}$ event by event \cite{antinori}. Based on 
current event generator predictions one expects that, even at LHC energies, 
all the higher  strangeness particles will be out of reach for an 
event-by-event study.

The inclusion of a fourth SVT layer based on Silicon Strip
Detectors, as was suggested by the French STAR collaborators, has been 
accepted by the collaboration as an official detector upgrade. It is estimated 
to improve the $\Lambda$ reconstruction efficiency by a factor three, and to 
raise 
the $\Xi$ and $\Omega$ efficiencies by about an order of magnitude. Although 
this will not lead to further event-by-event capabilities for strange baryons, 
it might be the necessary requirement towards successful $\Lambda\Lambda$ 
interferometry measurements due to the large increase of events in which 
at least a pair of $\Lambda$'s is reconstructed.

Because of the 
construction schedule of the SVT, the TPC will run without the SVT in RHIC 
year-1. In rapidity, the TPC extends slightly beyond the SVT whereas the 
inclusion of the SVT will improve the coverage at low transverse momenta.
Preliminary simulations show that with the TPC alone a neutral kaon 
measurement and a limited $\Lambda$ study are within reach in the first 
year of RHIC. The multi-strange baryon spectra and ratios will be
measured in the second year of RHIC, starting in the Fall of 2000.
At this time, the complete STAR strangeness program will be comparable to the 
SPS measurements of WA97 and NA49 combined.

\subsection{$\phi$-Meson reconstruction} 

It was suggested by several authors \cite{shuryak},\cite{brown}
that the decay of the $\phi$-meson in a baryon dense medium, could lead
to a signal for medium modifications in hadronic matter. Because of
its short life-time the $\phi$ might decay when off mass shell,
and therefore the $\phi$ mass peak as reconstructed from the decay 
products would be shifted. The $\phi$
decays into two main charged channels, namely the K$^{+}$K$^{-}$ channel 
and the e$^{+}$e$^{-}$ channel. Although the branching ratios of the two
channels are very different (49.1$\%$ for the kaon channel, 3$\times$10$^{-4}$
for the lepton channel) a high precision measurement of the $\phi$
mass peak, requiring a resolution better than the natural width
of the $\phi$ (= 4.4 MeV), is possible in both channels. The kinematic limit 
in the K$^{+}$K$^{-}$-channel is just 32.6 MeV below the $\phi$-mass and based on the 
lifetime of the $\phi$ only a fraction of the produced $\phi$'s will
decay in medium. Thus, in this decay branch, the effect will most likely 
only cause a widening of the mass peak.
A more reliable measurement therefore is the determination of a
relative change in the branching ratios by simultaneously measuring both 
charged decay channels. PHENIX and STAR have good capabilities for measuring 
the $\phi$ decay branching ratios. Fig.4 shows simulations of the PHENIX 
$\phi$ mass peak
in both channels. These studies were done for early stages in the PHENIX
data taking. The $\phi$ to K$^{+}$K$^{-}$ studies were done for 3 million 
central events (the estimated data volume that will be collected during the 
first few weeks of data taking at reduced luminosity) assuming only 1/2 of one 
PHENIX detector arm is instrumented. The e$^{+}$e$^{-}$ studies assume
the first full year of running at reduced luminosity with both
arms fully instrumented. 

\begin{figure}[hbtl]
\begin{center}
\mbox{
\epsfig{file=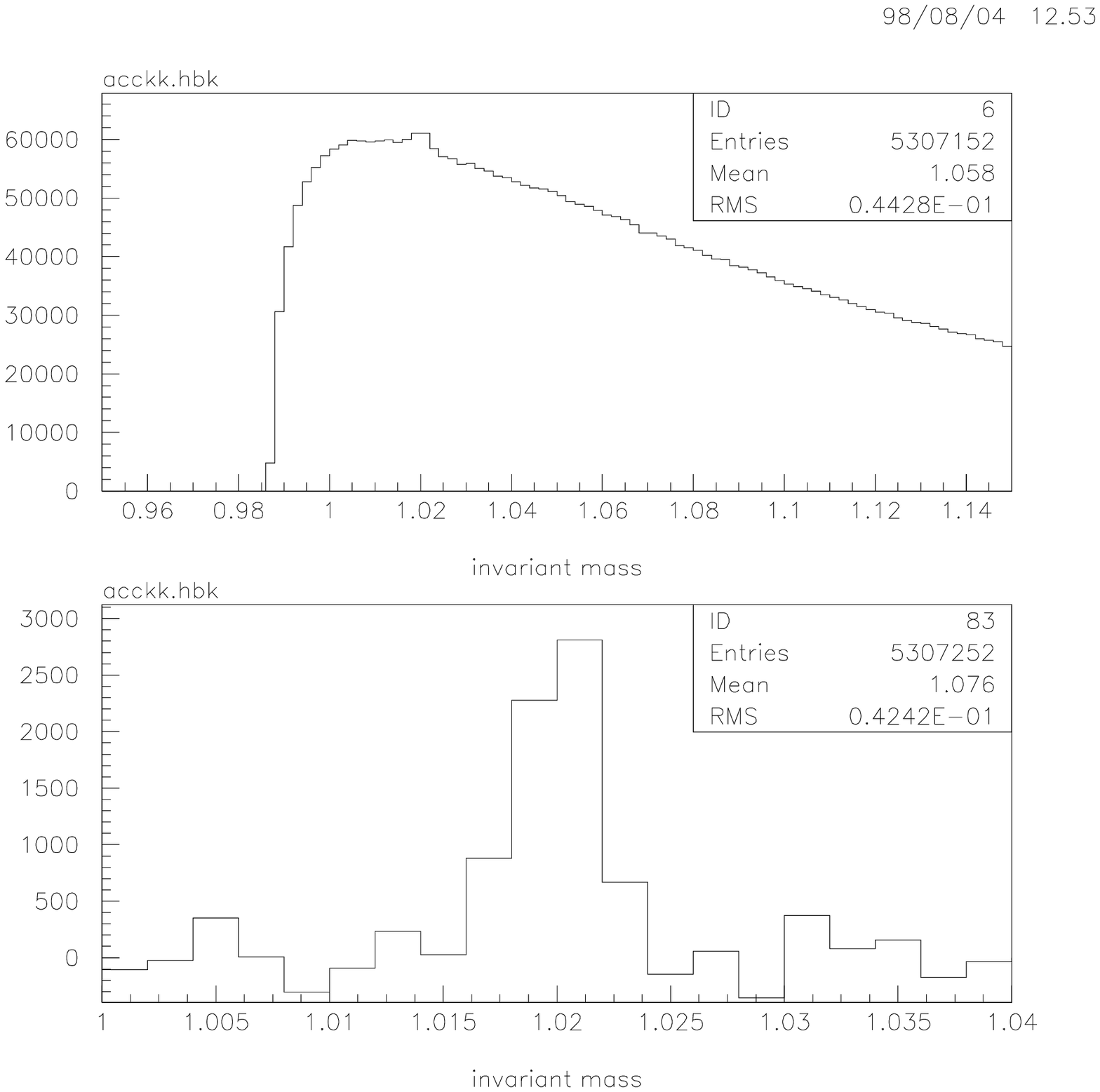,bbllx=50pt,bblly=155pt,bburx=530pt,bbury=660pt,width=5.6cm}
}
\mbox{
\epsfig{file=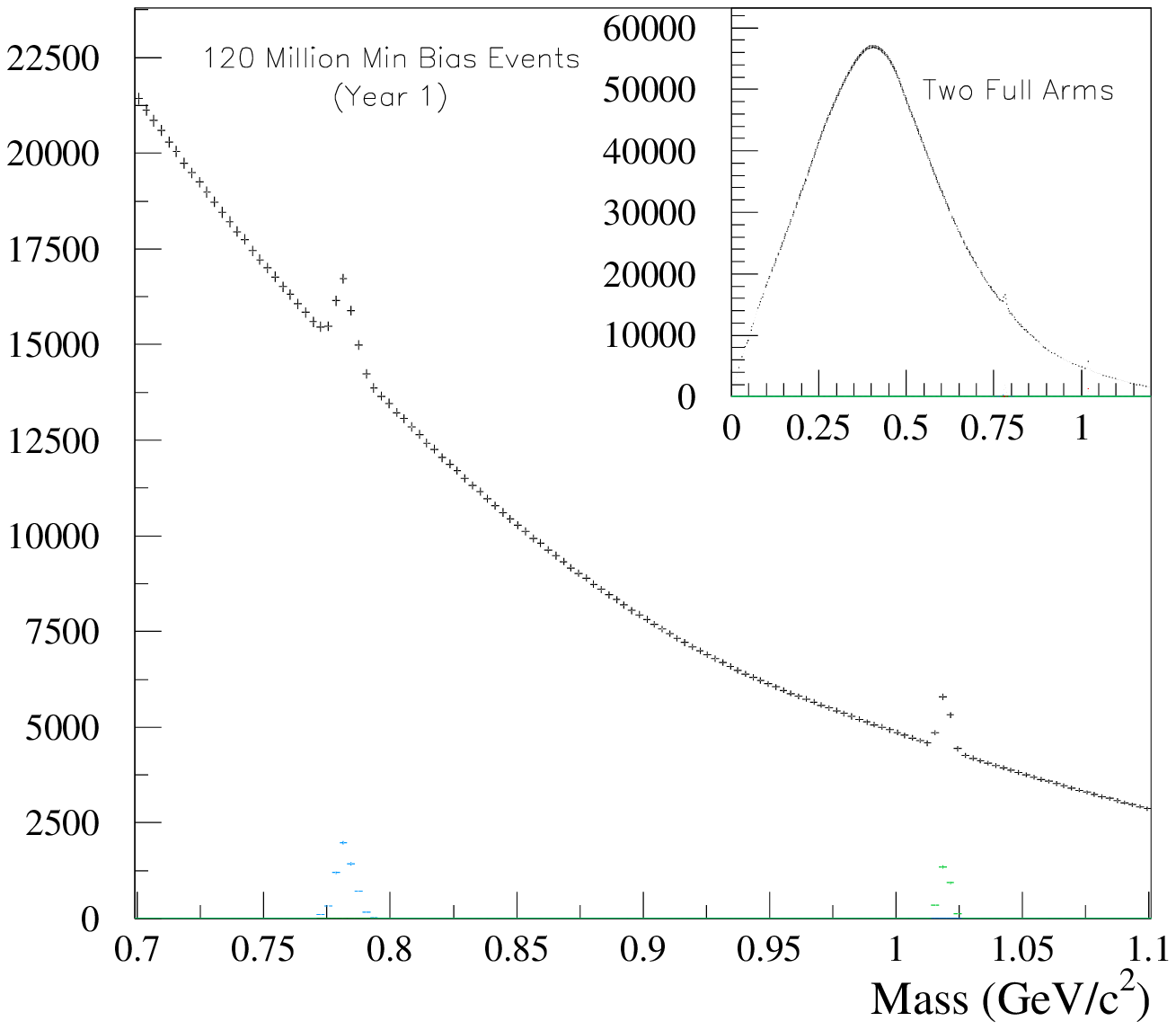,bbllx=40pt,bblly=40pt,bburx=420pt,bbury=400pt,width=6.6cm}
}
\end{center}
\caption{PHENIX reconstructed Invariant Mass spectrum a.) in the K$^{+}$K$^{-}$-channel
and b.) in the e$^{+}$e$^{-}$-channel}
\end{figure}

\subsection{Strange Quark Matter}

In most of the theoretical predictions for strange quark matter
formation in relativistic heavy ion collisions {\cite{carsten3}, the
main requirements are large baryon densities and a relatively small
bag pressure and energy density. Therefore most of the past experiments
were performed at AGS energies. Jack Sandweiss gave
a nice summary of the experimental status at this conference
\cite{jack1}. Based on these arguments, the probability of forming a
strangelet in the central rapidity region at RHIC seems low, but the fact
that the forward rapidity regions at RHIC are expected to have high
net baryon densities, might lead to an interesting phenomenon, namely the
possible formation of two separate phases of QGP in rapidity,
distinguishable by their respective baryonic content. In this scenario 
the probability of
forming strange quark matter is enhanced at forward rapidities.
The suggested detection methods are similar to the ones presently employed
by E864 (AGS) and NA52 (SPS) for long-lived strangelets and by E896 (AGS)
for short-lived strangelets. The STAR Silicon Detector Prototypes were used
in E896 to detect short lived H-Dibaryons via the $\Sigma^{-}$p-channel.
The analysis of those measurements is still ongoing. The detection
of the long-lived strangelets in the STAR tracking system would be 
based on the characteristic energy loss behavior of particles with an
unusual Z/A-ratio. A detailed study of these effects can be found in
\cite{french}.

\section{Discussion and Conclusions}

By studying the list of simulations shown in the previous section
it becomes obvious that the Strangeness Physics program proposed at RHIC is
very similar to the one at the AGS and SPS. Thus, the question remains,
whether the measurement of strangeness observables at RHIC will lead to
any new physics, or whether the conclusions deduced from strangeness
measurements at AGS and SPS are still applicable at the higher energies.
In other words, is strangeness still interesting at RHIC ? 
Based on my interpretation of the latest model calculations for the RHIC 
energy regime my answer would be an unequivocal 'Yes', but my arguments might
be different than expected.

During this conference most AGS and SPS measurements were
described with models assuming full thermal and chemical equilibration
\cite{sollfrank}, \cite{schneder}.
Both theory talks from Heinz \cite{heinz} and Becattini \cite{beccatini}
documented evidence for thermal and chemical equilibration, and in
particular Heinz pointed out the quantitative difference between chemical
equilibration and freeze-out as measured by particle ratios, and thermal
equilibration and freeze-out as measured by particle distributions. Also,
the effect of collective expansion is by now well parameterized and 
understood. Rafelski \cite{rafelski} showed again that certain particle
ratios, in particular those including strange anti-baryons \cite{stephans} and 
multi strange baryons \cite{wa97} can not be easily explained with
hadronic thermalization or even with string fragmentation, leading
to the conclusion that we are observing an onset of a new phenomenon
at AGS and SPS energies \cite{welke}, \cite{rafelski2}. This point is 
obviously much debated and
some thermal models \cite{pbm} as well as interacting string fragmentation 
models \cite{sorge},\cite{werner} have claimed that many of the higher 
particle ratios and distributions can be described with the simple assumption
of an interacting dense hadron gas.

For RHIC energies the situation is less uncertain because the expected energy
density simply requires a description based on perturbative QCD, at least
for the very early part of the interaction. Even more, most high energy
models agree that the heavy ion collision at RHIC will undergo three
distinct phases, namely the initial hard parton-parton scattering, then
a pre-equilibrium phase, and finally the hadronic equilibration that
seems to be well understood at lower energies. The first stage is well
described by perturbative QCD, the second stage is addressed by parton
cascade models \cite{vni},\cite{hijing-b}, and the third stage which 
includes the hadronization and
subsequent final state interactions has recently been modeled by combining
an initial parton cascade with either hydrodynamical or 
string fragmentation-like transport codes. This merging of the three phases 
of the interaction
into one continuous code seems the most complete way of describing the
RHIC interaction, and in the following I would like to focus on two
recent publications, namely the parton cascade-hydro model by Dumitru
and Rischke \cite{risch98}, and the combined VNI-parton cascade/HIJET-hadron 
code by Geiger and Longacre \cite{geiger}. Although
neither model should be taken too literally at this point, the interpretation
of some of the results allows us to shed some light on what to expect at
RHIC, especially for the production of strange baryons and mesons.

Fig.5 shows the relative contributions of each of the three stages to the 
final hadron distribution in a particular model \cite{geiger}.

\begin{figure}[htbl]
\begin{center}
\mbox{\epsfig{file=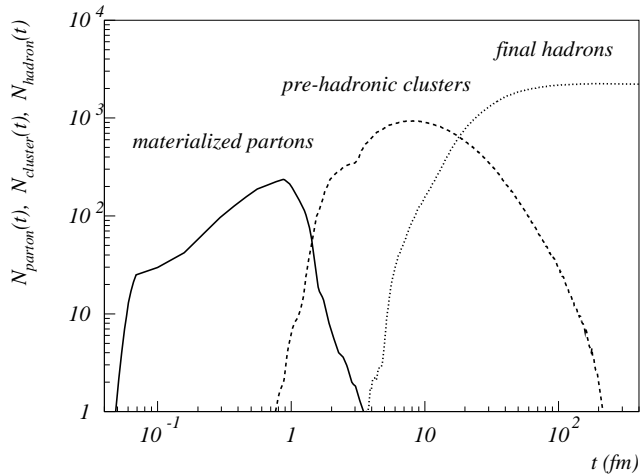,bbllx=20pt,bblly=200pt,bburx=550pt,bbury=600pt,width=9cm}}
\end{center}
\caption{Relative contributions from the various reaction stages to the 
final hadron distribution (from \cite{geiger}).}
\label{fig:acceptance}
\end{figure}

We assume, based on our present knowledge of the anticipated RHIC energy density
and the shape of the quark and gluon structure functions, that the initial 
hard scattering phase is gluon dominated. Therefore the main contribution to 
strangeness formation in the early part of the collision is due to gluon 
fusion, gluon decay and gluon scattering. Early work by Geiger \cite{geiger2}
predicted a very large strangeness enhancement based on a phase transition.
Generally, models that include a parton cascade greatly enhance the 
multiplicity of produced particles, in particular by increasing the number 
of produced pions.
The strange quarks will most likely not chemically equilibrate 
\cite{gyulassy} but the parton cascade might lead to fast local thermal
equilibration. The significant strangeness formation in the parton
cascade, mostly due to re-interaction of soft gluons has to be taken
into account when defining the initial condition for a subsequent
hadronic transport code.
Fig.6 shows a simulation by Dumitru and Rischke \cite{risch98} which 
documents the transverse momentum spectra for direct thermal hadrons
at RHIC assuming a parton cascade followed by a hydrodynamical transport.
Obviously the shape of the spectra can be analyzed to determine collective
dynamics at RHIC. m$_{T}$-scaling is only valid in a small transverse
momentum range from 1.5-2.5 GeV/c, whereas the low p$_{T}$ and the high p$_{T}$
part of the spectrum show significant deviation from the scaling, and thus
provide important collective information.

\begin{figure}[htbl]
\begin{center}
\mbox{\epsfig{file=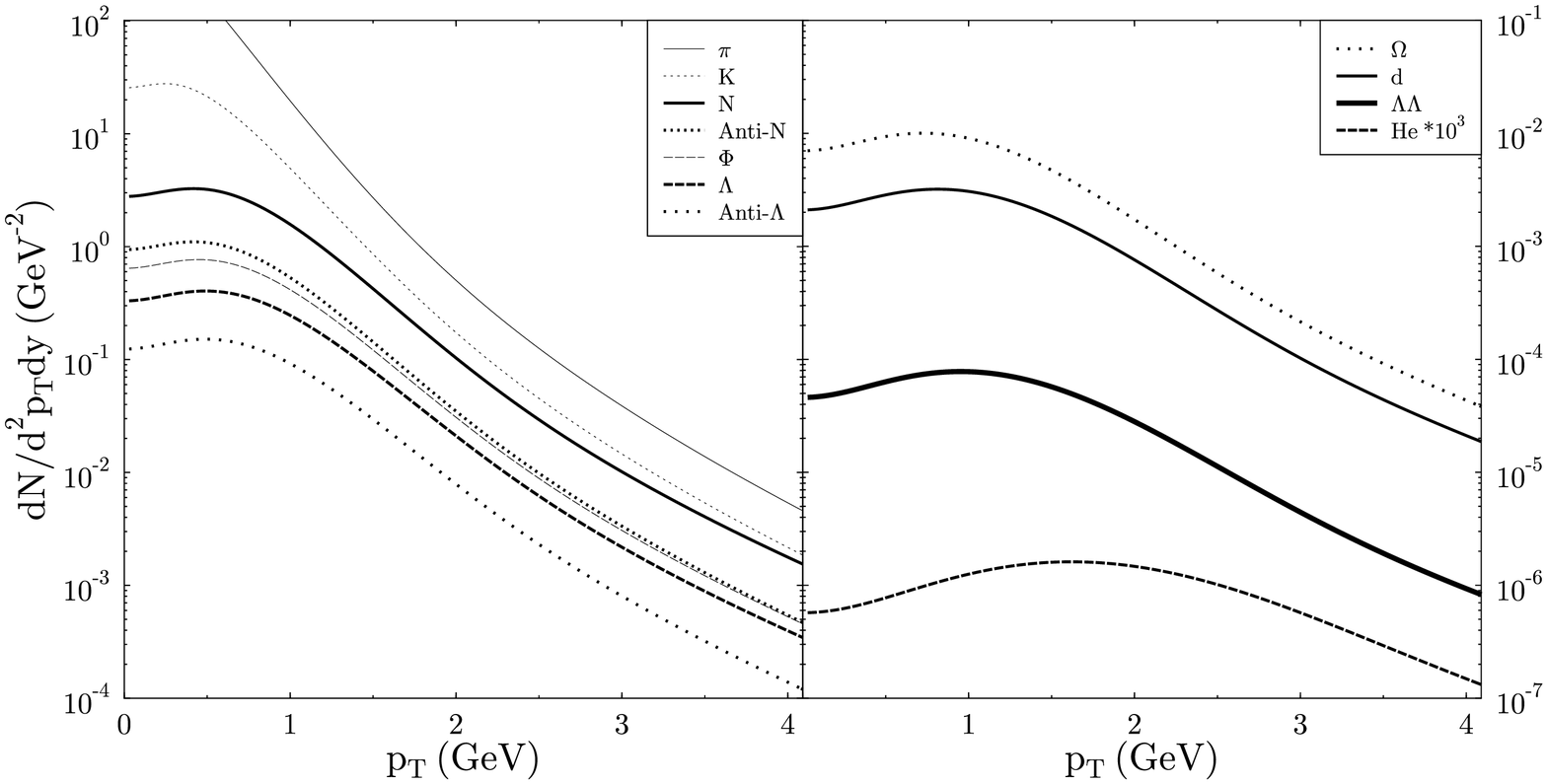,bbllx=0pt,bblly=0pt,bburx=720pt,bbury=400pt,width=12cm}}
\end{center}
\caption{Transverse momentum spectra for direct thermal hadrons at RHIC
energies from a hydrodynamical simulation assuming freeze-out at T = 130 MeV
(from \cite{risch98}).}
\label{fig:acceptance}
\end{figure}

Another interesting feature of the hadronic transport after a parton
cascade is shown in Fig.7. Here we compare the rapidity 
distributions for non-strange baryons and anti-baryons with and without a 
hadronic 'afterburner' following particle generation via parton cascade 
\cite{geiger}. The net baryon density close to mid-rapidity is not zero,
but it is small. However, there is significant baryon density in forward
direction. The hadron cascade requires non-strange quarks
for its $\Lambda$ formation, whereas the $\overline{\Lambda}$ formation requires
non-strange anti-quarks. Thus, the strange baryon and anti-baryon distributions
follow the distributions of the protons and anti-protons. This leads to
distinctly 
different rapidity distributions for the $\Lambda$ and the $\overline{\Lambda}$. 
The $\Lambda$, in this model, is peaked at more forward rapidities 
and shows a distinct 'dip' at mid-rapidity, whereas the $\overline{\Lambda}$ 
is peaked at mid-rapidity as expected. 

\begin{figure}[hbtl]
\begin{center}
\mbox{
\epsfig{file=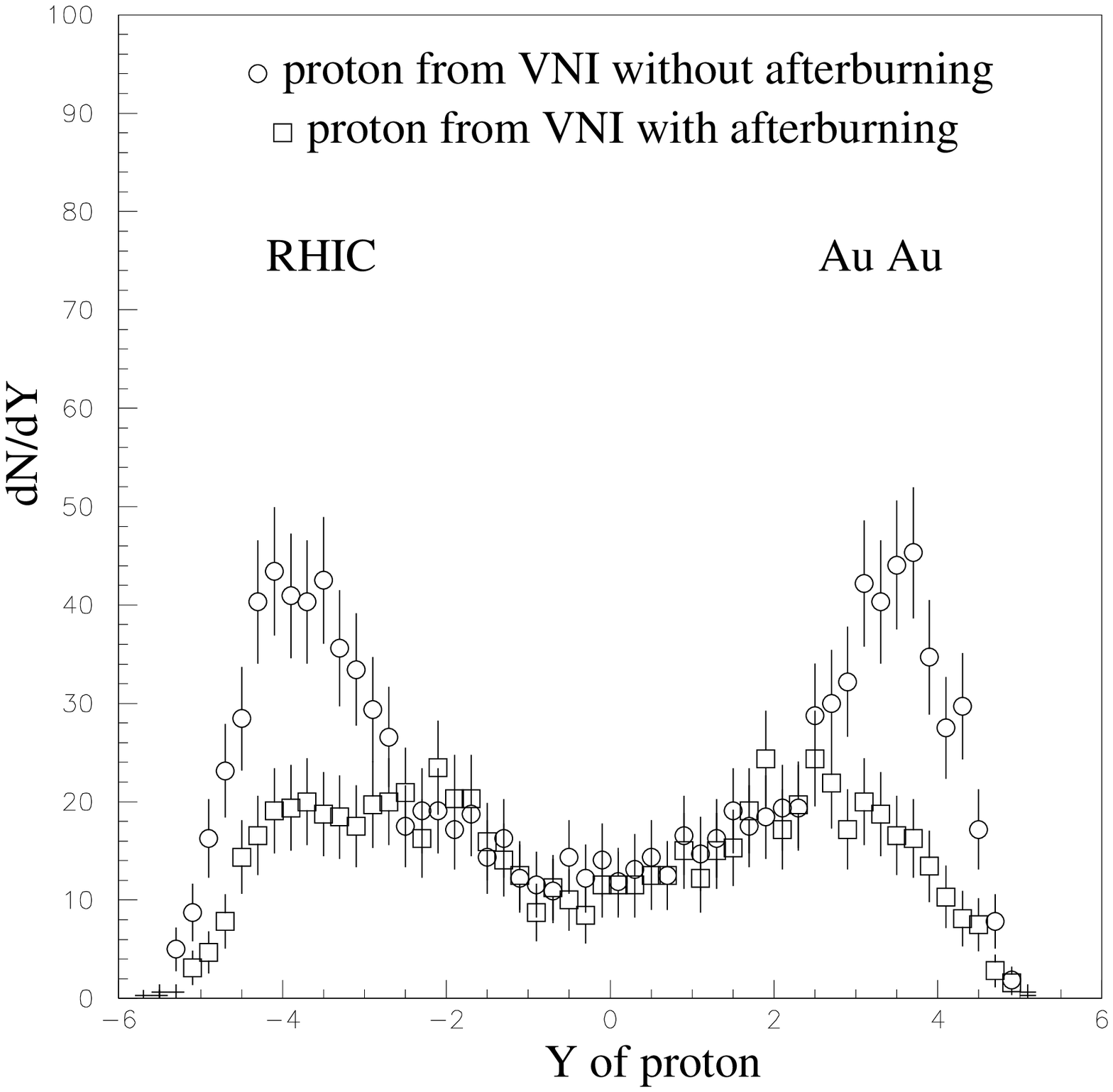,bbllx=50pt,bblly=155pt,bburx=520pt,bbury=645pt,width=7.cm}
}
\mbox{
\epsfig{file=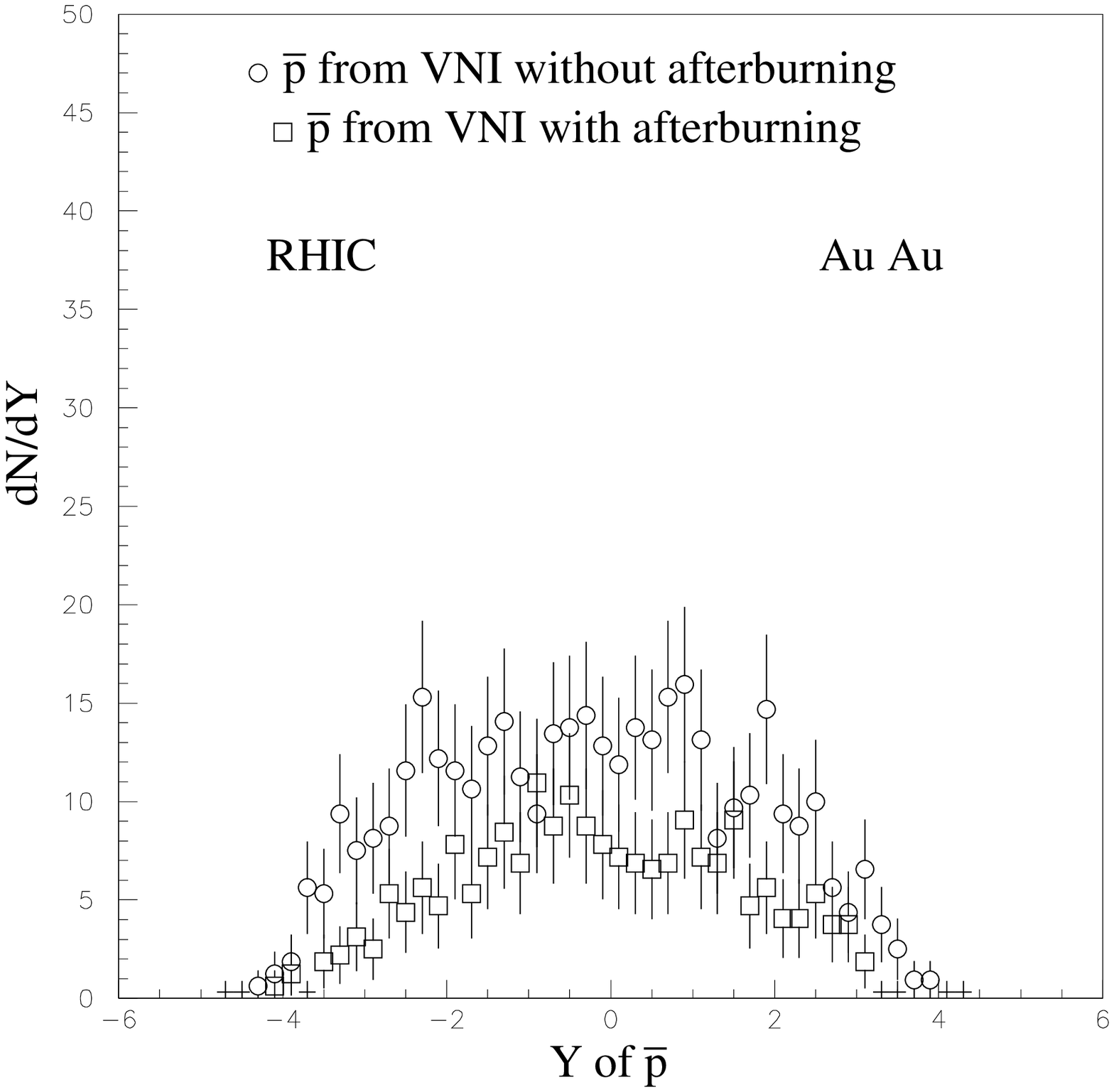,bbllx=50pt,bblly=155pt,bburx=520pt,bbury=645pt,width=7.cm}
}
\end{center}
\caption{Rapidity distributions of p, $\overline{p}$ based on the 
VNI parton cascade with and without hadronic 'afterburner' in Au+Au collisions 
at RHIC energies (from \cite{geiger}).}
\end{figure}

These simulations are very model-dependent and should be taken with a grain 
of salt, but they might point us at the measurements we have to 
analyze to shed some light on the strangeness production mechanisms in RHIC 
collisions.

In summary, I would reiterate on my conclusion that strangeness is still
interesting at RHIC. Because of the nature of the interactions at RHIC,
the emphasis of strangeness measurements might shift away from
the question of phase transition detection through chemically enhanced
and equilibrated signals to the question of whether the different interaction
phases in a collision at RHIC energies leave remnants in the final measurement. 
As an example, we know that any parton cascade model requires a transition, 
the question is whether this transition can be measured in quantities like
the strangeness yield which seems to combine several production mechanisms
in a single observable.
As I tried to show in the few theoretical predictions, basic strangeness
measurements, like yield, ratios, rapidity and transverse momentum
distribution are an important piece in the big puzzle of plasma formation.
They will be most useful in correlation with many other hadronic and
leptonic measurements at RHIC. Most likely, no single measurement will
be sufficient to define the phase transition. In this sense one should
view strange baryons as a.) rare signals, b.) heavy baryons, and c.)
produced particles. In these categories, strangeness will make its
contribution, together with many other observables, towards detecting the 
QGP.

\section{Acknowledgements}

I would like to thank Adrian Dumitru, 
Klaus Kinder-Geiger, Gerd Kunde, Dirk Rischke, Richard Seto, Robert Willson, 
and Ken Wilson for useful discussions and their contributions. 
Helen Caines, John Harris, and Claude Pruneau are gratefully acknowledged for
reading the manuscript.

\end{document}